\documentclass[conference]{IEEEtran}
\pdfoutput=1 
\IEEEoverridecommandlockouts
\usepackage{algorithm} 		
\usepackage{algpseudocode}
\usepackage{cite}
\usepackage{amsmath,amssymb,amsfonts}
\usepackage{float}
\usepackage{graphicx}
\usepackage{textcomp}
\usepackage{xcolor}
\usepackage{color}
\usepackage{stfloats}
\usepackage{subcaption}
\usepackage{array}
\usepackage{multirow}
\usepackage{caption}
\usepackage{subcaption}
\usepackage{booktabs}
\usepackage{tabularx}
\usepackage[colorlinks=true,linkcolor=black,citecolor=colorlinks=true,linkcolor=black,citecolor=blue,urlcolor=blue,urlcolor=blue]{hyperref}

\begin{document}

\title{A Novel Low-Complexity Dual-Domain Expectation Propagation Detection Aided AFDM for Future Communications}
\author{
	Qin~Yi$^\dag$, Ping~Yang$^\dag$, Zilong~Liu$^*$, Zeping~Sui$^*$, Yue~Xiao$^\dag$, and Gang~Wu$^\dag$ \\ 
	\normalsize 
      $^\dag$National Key Laboratory of Wireless Communications, \\University of Electronic Science and Technology of China, Chengdu, 611731, China\\
     $^*$School of Computer Science and Electronics Engineering, University of Essex, Colchester, CO4 3SQ, United Kingdom\\
   Email: yang.ping@uestc.edu.cn
\vspace{-2em}
\thanks{This work is supported by the National Key R\&D Program of China under Grant 2025YFE0100900, and the National Key Science and Technology Major Project of China (2025ZD1302000).}}

\maketitle

\begin{abstract}
This paper presents a dual-domain low-complexity expectation propagation (EP) detection framework for affine frequency division multiplexing (AFDM) systems. By analyzing the structural properties of the effective channel matrices in both the time and affine frequency (AF) domains, our key observation is the domain-specific quasi-banded sparsity patterns, including AF-domain sparsity under frequency-selective channels and time-domain sparsity under doubly-selective channels. Based on these observations, we develop an AF-domain EP (EP-AF) detector for frequency-selective channels and a time-domain EP (EP-T) detector for doubly-selective channels, respectively.
By performing iterative inference in the time domain using the Gaussian approximation, the proposed EP-T detector avoids inverting the dense channel matrix in the AF domain. Furthermore, the proposed EP-AF and EP-T detectors leverage the aforementioned quasi-banded sparsity of the AF domain and time domain channel matrices, respectively, to reduce the complexity of matrix inversion from cubic to linear order.
Simulation results demonstrate that the proposed low-complexity EP-AF detector achieves nearly identical error rate performance to its conventional counterpart, while the proposed low-complexity EP-T detector offers an attractive trade-off between detection performance and complexity.
\end{abstract}


\section{Introduction}
Reliable data transmission in high-mobility scenarios, such as high-speed railways, unmanned aerial vehicles, and satellite links, is a critical requirement for the next generation wireless communication systems \cite{HighMob}. However, such environments typically incur severe Doppler spread, which induces substantial inter-carrier interference in conventional orthogonal frequency division multiplexing (OFDM) systems, resulting in significant degradation in bit error rate (BER) performance \cite{AFDM_SCMA}. To overcome this limitation, affine frequency division multiplexing (AFDM) has been proposed as a Doppler-resilient waveform, offering reliable communication performance in high-mobility channels \cite{AFDM}. In AFDM, information symbols in the affine frequency (AF) domain are modulated via the inverse discrete affine Fourier transform (IDAFT) onto a set of orthogonal chirp subcarriers that span the entire time-frequency domain. By carefully tuning the chirp rate according to the maximum Doppler spread, AFDM enables a separable channel representation in the AF domain and achieves full diversity \cite{AFDM_Para}. Furthermore, the DAFT can be efficiently implemented using the fast Fourier transform, making AFDM both computationally efficient and highly backward-compatible with existing OFDM-based transceivers \cite{NonAFDM}.

It is worth noting that in order to achieve the benefits of AFDM, effective detector design plays a critical role. Recently, several detection algorithms have been proposed to enhance the performance of AFDM systems. Specifically, a low-complexity minimum mean square error (MMSE) detector and a weighted maximum ratio combining (MRC)-based decision feedback equalizer were proposed in \cite{AFDM_MRC}, both of which exploit the sparsity of the effective channel matrix in the AF domain to reduce complexity. 
Moreover, based on a sparse factor graph and the Gaussian approximation of interference, a low-complexity message passing (MP) detector was introduced in \cite{AFDM_MP}. However, the aforementioned low-complexity MMSE, MRC, and MP detectors suffer from significant performance degradation in doubly-selective channels, where fractional Doppler effects cause the effective channel matrix in the AF domain to become dense and lose its sparsity. 

To overcome the limitations of existing detectors in doubly-selective channels and fully exploit the domain-specific sparsity of channel matrices, this paper proposes a dual-domain low-complexity expectation propagation (EP) detection framework for AFDM systems, which adaptively selects the inference domain (AF or time) according to the underlying channel characteristics. The main contributions of this work are summarized as follows:
\begin{itemize}
\item{{\bf Dual-domain EP detection tailored to channel conditions:} For frequency-selective channels, our key observation is that the effective channel matrix in the AF domain exhibits a quasi-banded sparse structure. Based on this observation, we propose an AF-domain EP (EP-AF) detector that exploits this property for efficient inference.
For doubly-selective channels, in a similar way, we develop a time-domain EP (EP-T) detector, which leverages the quasi-banded structure of the time domain channel matrix and approximates the posterior distribution of time domain symbols as a Gaussian distribution to progressively refine symbol detection.}
\item{{\bf Sparsity-aware matrix inversion via lower-upper (LU)  decomposition:} By leveraging the quasi-banded sparsity of the effective channel matrices in their respective domains, both EP-AF and EP-T detectors employ LU factorization to significantly reduce the complexity of matrix inversion from cubic to linear order.}
\item {{\bf Performance-complexity trade-off:} Numerical results demonstrate that the proposed low-complexity EP-AF detector attains performance nearly identical to that of its conventional counterpart while substantially reducing computational complexity. Meanwhile,  the proposed low-complexity EP-T detector is capable of achieving a favorable trade-off between BER performance and complexity.}
\end{itemize}

\captionsetup[figure]{singlelinecheck=off}
\begin{figure*}
    \centering
    \includegraphics[width=140mm]{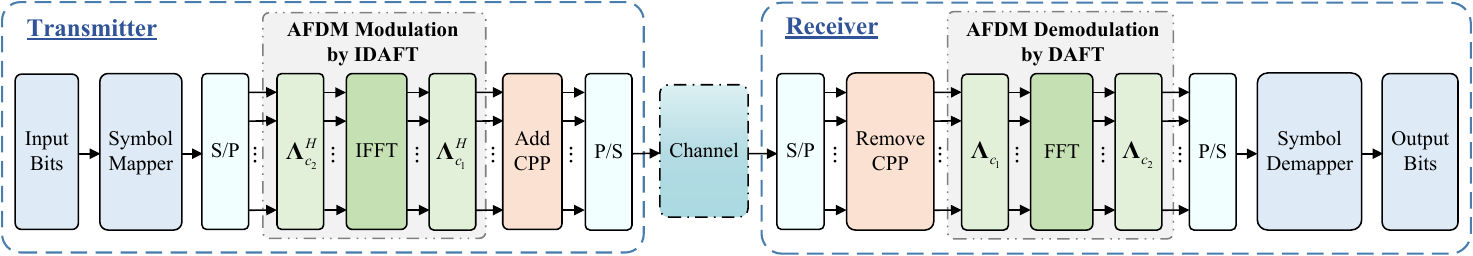}
    \captionsetup{subrefformat=parens} 
    \caption[left]{Block diagram of the AFDM transceiver system.}
    \label{fig:SystemModel}
    \vspace{-2em}
\end{figure*}
\captionsetup[figure]{singlelinecheck=on}

\section{System Model}
The transceiver architecture of the AFDM system is illustrated in  Fig.~\ref{fig:SystemModel}. Let $\bar {\mathbf{x}}=\left[ {\bar{x}[0],\bar{x}[1], \ldots ,\bar{x}[N - 1]} \right] \in {\mathbb{C}^{{N}\times 1}}$ denote the vector of information symbols in the AF domain, where $N$ represents the number of chirp subcarriers. After the serial-to-parallel operation, the AF domain symbol vector is transformed into the time domain via the IDAFT, given by
\begin{align}
\label{eq:Modu_Sym}
    \bar{s}[n]=\sum\limits_{m=0}^{N-1} \bar{x}[m] \varphi_n(m),  n=0,\ldots, N-1,
\end{align} 
where $\varphi_n(m)= \frac{1}{\sqrt{N}} e^{\imath 2\pi (c_1 n^2 + c_2 m^2 + \frac{n m}{N})}$ denotes the orthogonal chirp basis function of the IDAFT, while $c_1$ and $c_2$ are two chirp parameters of the AFDM. In matrix form, \eqref{eq:Modu_Sym} can be written as $\bar{{\bf{s}} }= {{\bf{A}}^H}\bar{{\bf{x}}}$,
where $\mathbf{A}=\boldsymbol{\Lambda}_{c_2} \mathbf{F} \boldsymbol{\Lambda}_{c_1} $ is the DAFT matrix, $\mathbf{\Lambda}_{c} = \text{diag}\left(e^{-\imath 2\pi c n^2} , \, n = 0, \ldots, N-1 \right)$ is a diagonal matrix, and $\mathbf{F}$ denotes the discrete Fourier transform matrix with entries $ \frac{1}{\sqrt{N}} e^{-\frac{\imath2\pi m n}{N}}$.
To mitigate inter-symbol interference introduced by multipath propagation, an $N_{\mathrm{CPP}}$-length chirp-periodic prefix (CPP) is appended to the beginning of each transmitted time-domain signal.

The discrete-time impulse response of a doubly-selective channel with ${P}$ propagation paths is given by
\begin{align}\label{eq:channel}
g_n(l) = \sum_{i=1}^{{P}} h_i \, e^{-\imath \frac{2\pi}{N} \nu_i n} \, \delta(l - l_i),
\end{align}
where $h_i$, $\nu_i\in [-\nu_{\max}, \nu_{\max}]$, and $l_i\in [0, l_{\max}]$ represent the complex path gain, normalized Doppler shift, and delay shift of the $i$-th path, respectively, and $\delta(\cdot)$ denotes the Dirac delta function.
The maximum normalized Doppler shift is represented by $\nu_{\max}$, and the maximum delay spread is given by $l_{\max} = \max(l_i)$. We define \(\nu_i = \alpha_i + \beta_i\), 
where $\alpha_i \in [-\alpha_{\max}, \alpha_{\max}]$ and $\beta_i \in \left[-\frac{1}{2}, \frac{1}{2}\right)$ denote the integer and fractional components of $\nu_i$, respectively.

Accordingly, the received signal in the discrete-time domain is expressed as
\begin{align}\label{eq:RxSignal_TD}
  \bar{r}[n] = \sum_{l=0}^\infty \bar{s}[n-l] g_n(l) + \bar{w}_\mathrm{T}[n],
\end{align}
where $w_\mathrm{T}[n]\sim \mathcal{CN}(0, \sigma_{\bar w}^2)$ is the additive Gaussian noise, and $\sigma_{\bar w}^2$ denotes the noise variance. After removing the CPP, \eqref{eq:RxSignal_TD}
can be written in matrix form as
\begin{align}\label{eq:RxSym_TD}
\bar{\mathbf{r}}  = \bar{\mathbf{H}}_\mathrm{T} \bar{\mathbf{s}} + \bar{\mathbf{w}}_\mathrm{T},
\end{align} 
where $\bar{\mathbf{H}}_\mathrm{T} = \sum_{i=1}^{{P}} h_i \boldsymbol{\Gamma}_{\mathrm{CPP}_i} {\mathbf{\Delta}}_{\nu_i} \boldsymbol{\Pi}^{l_i}$ denotes the time domain channel matrix. Here, $\boldsymbol{\Pi}$ represents the forward cyclic-shift matrix,
${\mathbf{\Delta}}_{\nu_i} = \operatorname{diag}\left(e^{-\imath \frac{2 \pi}{N} \nu_i n}, n = 0, 1, \ldots, N-1\right)$, and $\boldsymbol{\Gamma}_{\mathrm{CPP}_i}$ is a diagonal matrix associated with the CPP \cite{AFDM_GSM,MM_AFDM}.

By applying the DAFT, the received signal in the AF domain can be expressed as 
\begin{align}\label{eq:RxSignal_AFD}
\bar{\mathbf{y}} = {\mathbf A} \bar{\mathbf{r}}  =\bar{\mathbf H}_\mathrm{AF} \bar{{\bf{x}}} + {\bar{\mathbf{w}}}_\mathrm{AF},
\end{align} 
 where $\bar{\mathbf H}_\mathrm{AF}={\mathbf A}\bar{\mathbf{H}}_\mathrm{T} {\mathbf A}^H$ denotes the effective channel matrix in the AF domain and ${\bar{\mathbf{w}}}_\mathrm{AF} = {\mathbf A}\bar{\mathbf{w}}_\mathrm{T}$ represents the noise vector. Since $\mathbf{A}$ is a unitary matrix, $\bar{\mathbf{w}}_\mathrm{AF}$ retains the same statistical properties as $\bar{\mathbf{w}}_\mathrm{T}$.

\section{Low-Complexity EP Detection for AFDM}
In this section, we first analyze the sparsity patterns of the channel matrices in both the time and AF domains. Based on these structural properties, a low-complexity EP-AF detector is proposed for frequency-selective channels by leveraging the quasi-banded sparsity of the AF domain channel matrix. For doubly-selective channels, we further develop a low-complexity EP-T detector that
exploits the sparsity of the time domain channel matrix. It approximates the posterior distribution of the time domain symbols as a Gaussian distribution and incorporates iterative information exchange between the time and AF domains to progressively refine symbol estimates under AF domain constellation constraints. Finally, a detailed complexity analysis is presented for both proposed detectors.


\subsection{Sparsity Analysis of Channel Matrices in the Time and AF Domains}\label{Sec:ChannelSpars}

In Fig.~\ref{fig:H_vmax0}, we illustrate the sparsity structure of the effective channel matrix $\bar{\mathbf{H}}_{\mathrm{AF}}$ and its Gram matrix $\mathbf{\Phi}_{\mathrm{AF}} = \bar{\mathbf{H}}_{\mathrm{AF}}^H \bar{\mathbf{H}}_{\mathrm{AF}}$ under frequency-selective channels. We can observe that $\bar{\mathbf{H}}_{\mathrm{AF}}$ exhibits a quasi-banded sparse structure with a bandwidth of $l_{\max} + 1$, while $\mathbf{\Phi}_{\mathrm{AF}}$ is also a banded matrix with a bandwidth of $ 2l_{\max} + 1$.  

\begin{figure}
    \centering
    \begin{subfigure}{0.1\textwidth} 
        \centering
        \includegraphics[width=\linewidth]{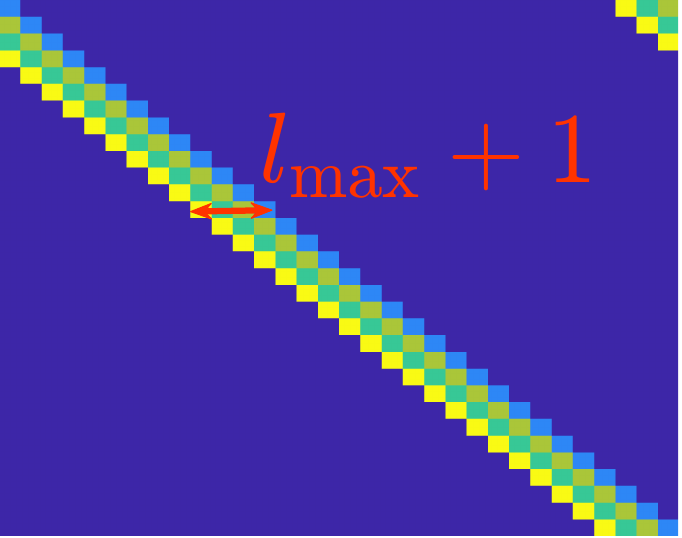}
        \subcaption{$\bar{\mathbf{H}}_\mathrm{T}$}
        \label{fig:vmax1Htx}
    \end{subfigure}
    \hspace*{0.1em} 
    \begin{subfigure}{0.1\textwidth}
        \centering
        \includegraphics[width=\linewidth]{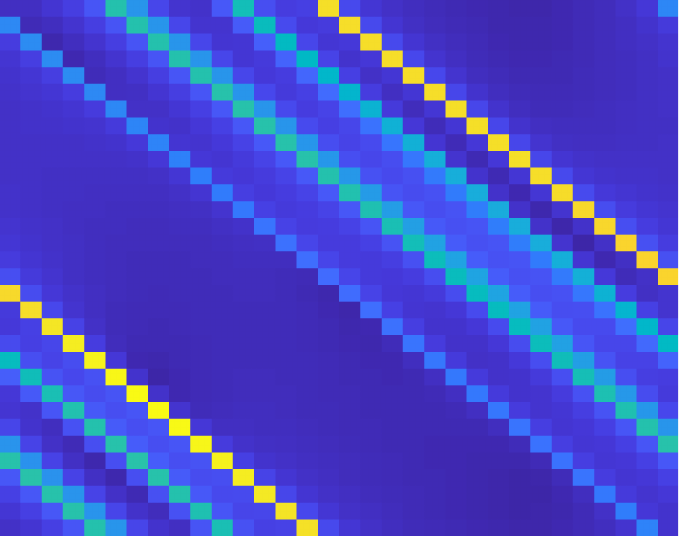}
        \subcaption{$\bar{\mathbf{H}}_\mathrm{AF}$}
        \label{fig:vmax1Heff}
    \end{subfigure}  
        \hspace*{0.1em} 
        \begin{subfigure}{0.1\textwidth}
        \centering
        \includegraphics[width=\linewidth]{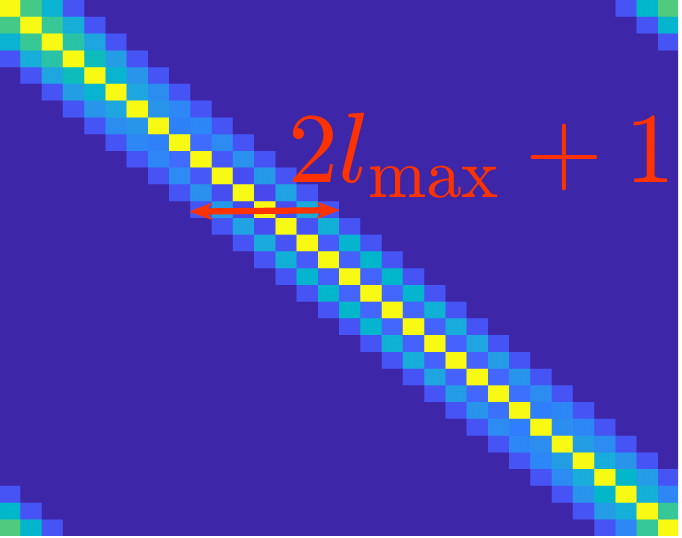}
        \subcaption{${\bf{\Phi }}_\mathrm{T}$}
        \label{fig:vmax1Fai_tx}
    \end{subfigure}
        \hspace*{0.1em} 
        \begin{subfigure}{0.1\textwidth}
        \centering
        \includegraphics[width=\linewidth]{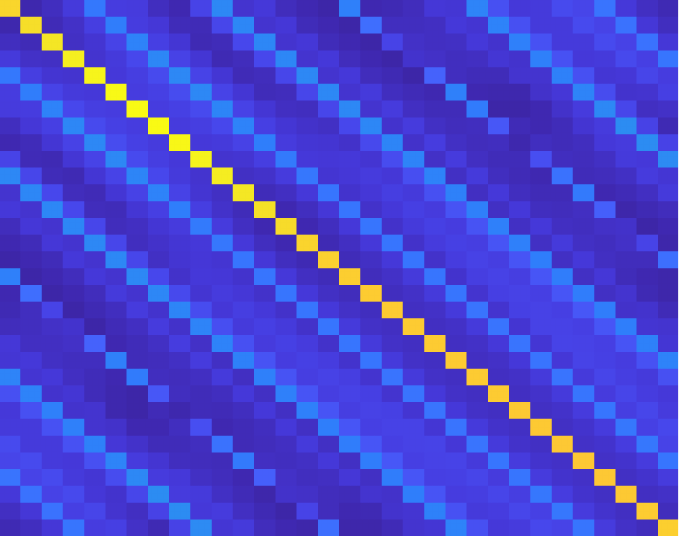}
        \subcaption{${\bf{\Phi }}_\mathrm{AF}$}
        \label{fig:vmax1Fai_eff}
    \end{subfigure}
    \captionsetup{subrefformat=parens} %
    \caption{Illustration of matrix sparsity in the time and AF domains with $N=32$, ${P}=4$, and $\nu_\mathrm{max} = 1$ under doubly-selective channels.}
\label{fig:H_vmax1}
\vspace{-2em}
\end{figure}

Fig.~\ref{fig:H_vmax1} presents a comparison of matrix sparsity in the time and AF domains under doubly-selective channels. It can be seen that the time domain channel matrix $\bar{\mathbf{H}}_{\mathrm{T}}$ shows a clear quasi-banded sparse structure with a bandwidth of $l_{\max} + 1$, and its Gram matrix $\mathbf{\Phi}_{\mathrm{T}} = \bar{\mathbf{H}}_{\mathrm{T}}^H \bar{\mathbf{H}}_{\mathrm{T}}$ retains this structure with a bandwidth of $2l_{\max} + 1$. In contrast, the AF domain matrix $\bar{\mathbf{H}}_{\mathrm{AF}}$ appears much denser due to the presence of fractional Doppler shifts, and its Gram matrix $\mathbf{\Phi}_{\mathrm{AF}}$ is nearly fully populated.

Motivated by these structural observations, we propose two low-complexity EP algorithms designed for different channel conditions: EP-AF for frequency-selective channels and EP-T for doubly-selective channels. The details of both algorithms are presented in the following subsections.

\subsection{EP-AF Detection for Frequency-Selective Channels}\label{Sec:EP-AF}
The EP-AF detector approximates the posterior distribution of AF domain symbols by replacing the discrete prior with a Gaussian distribution, and iteratively applies the moment matching method to refine the approximation \cite{OTFS_EP}.

\begin{figure*}
    \centering
    \includegraphics[width=130mm]{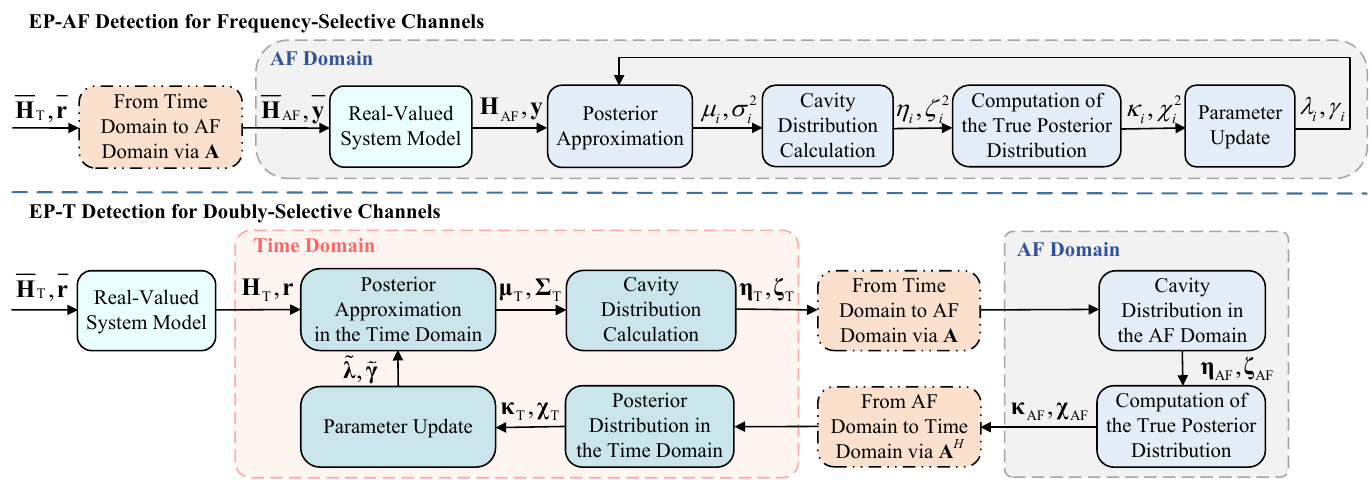}
    \captionsetup{subrefformat=parens} 
    \caption[left]{Block diagram of the proposed EP-AF detector for frequency-selective channels and EP-T detector for doubly-selective channels.}
    \label{fig:EP_Process}
    \vspace{-2em}
\end{figure*}

The complex-valued system model \eqref{eq:RxSignal_AFD} is converted into the real-valued one as 
\begin{align}\label{eq:RxSignal_AFD_Real}
{\bf{y}} = {{\bf{H}}_{{\rm{AF}}}}{\bf{x}} + {{\bf{w}}_{{\rm{AF}}}}
\end{align}
where 
\vspace{-0.5em}
\begin{align}
{\bf{y}}& = {\left[ {\begin{array}{*{20}{c}}
{\Re {{\{ {\bf{\bar y}}\} }^T}},{\Im {{\{ {\bf{\bar y}}\} }^T}}
\end{array}} \right]^T}\in \mathbb{R}^{2N}, \\
{\bf{x}} &= {\left[ {\begin{array}{*{20}{c}}
{\Re {{\{ {\bf{\bar x}}\} }^T}},{\Im {{\{ {\bf{\bar x}}\} }^T}}
\end{array}} \right]^T}\in \mathbb{R}^{2N},
\end{align}
\begin{align}\label{eq:Heff_Real}
{{\bf{H}}_{{\rm{AF}}}} &= \left[ {\begin{array}{*{20}{c}}
{\Re \{ {{{\bf{\bar H}}}_{{\rm{AF}}}}\} }&{ - \Im \{ {{{\bf{\bar H}}}_{{\rm{AF}}}}\} }\\
{\Im \{ {{{\bf{\bar H}}}_{{\rm{AF}}}}\} }&{\Re \{ {{{\bf{\bar H}}}_{{\rm{AF}}}}\} }
\end{array}} \right]\in \mathbb{R}^{2N\times2N},\\
{{\bf{w}}_{{\rm{AF}}}} &= {\left[ {\begin{array}{*{20}{c}}
{\Re {{\{ {{{\bf{\bar w}}}_{{\rm{AF}}}}\} }^T}},{\Im {{\{ {{{\bf{\bar w}}}_{{\rm{AF}}}}\} }^T}}
\end{array}} \right]^T}\in \mathbb{R}^{2N}.
\end{align}
According to \eqref{eq:RxSignal_AFD_Real}, the posterior probability of the transmitted symbol vector $\mathbf{x}$ can be expressed as
\begin{align}
p({\bf{x}}|{\bf{y}}) 
\propto f({\bf{x}})\triangleq{\cal N}({\bf{y}};{{\bf{H}}_{{\rm{AF}}}}{\bf{x}},\sigma _w^2{{\bf{I}}_{2N}})\prod\limits_{i = 1}^{2N} {{{\mathbb{I}_{{x_i} \in {\cal A}}}}}, 
\end{align}
where $\mathbb{I}_{x_i \in \mathcal{A}}$ denotes the indicator function that constrains the value of ${x_i \in \mathcal{A}}$, $\mathcal{A}$ is the modulation alphabet, and $\sigma_w^2=\sigma_{\bar w}^2/2$. 
As illustrated in Fig.~\ref{fig:EP_Process}, the main steps of the EP-AF algorithm are detailed as follows:


\begin{figure}
    \centering
    \begin{subfigure}{0.12\textwidth}
        \centering
        \includegraphics[width=\linewidth]{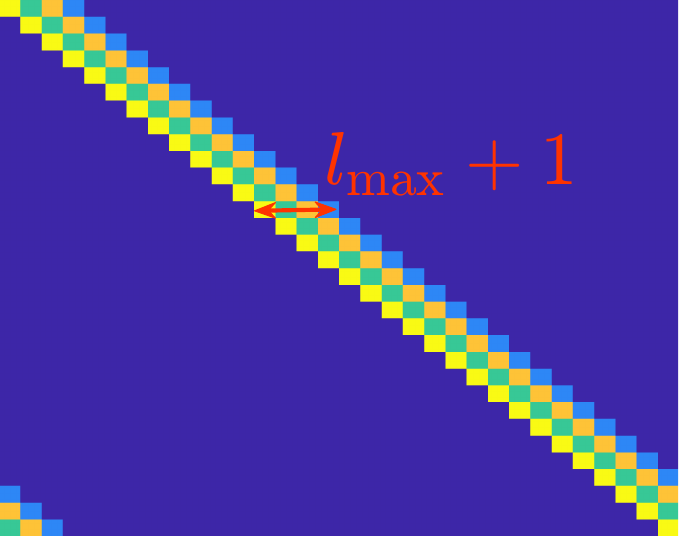}        \subcaption{$\bar{\mathbf{H}}_\mathrm{AF}$}
        \label{fig:vmax0Heff}
    \end{subfigure}  
        \hspace*{2em} 
        \begin{subfigure}{0.12\textwidth}
        \centering
        \includegraphics[width=\linewidth]{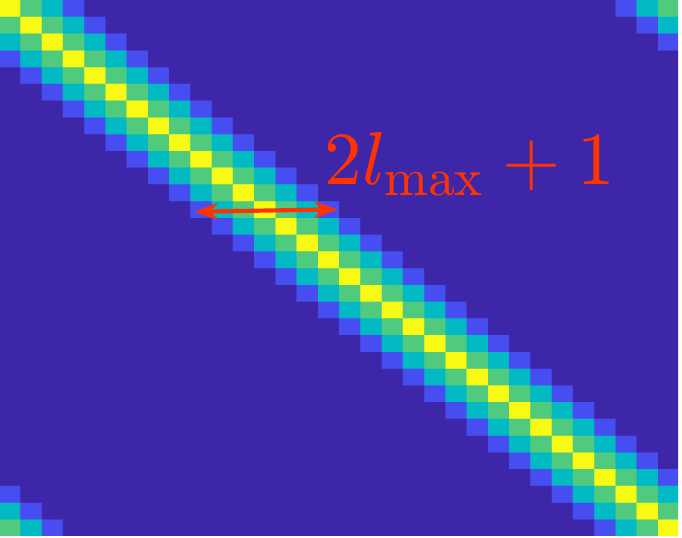}       \subcaption{${\bf{\Phi}}_\mathrm{AF}$}
        \label{fig:vmax0Fai_eff}
    \end{subfigure}
    \captionsetup{subrefformat=parens} %
    \caption{Illustration of matrix sparsity in the AF domain with $N=32$ and ${P}=4$ under frequency-selective channels.}
\label{fig:H_vmax0}
\vspace{-1em}
\end{figure}

\captionsetup[figure]{singlelinecheck=off}
\captionsetup[figure]{singlelinecheck=on}

\subsubsection{Posterior Approximation}
To facilitate tractable inference, each non-Gaussian factor in $p({\mathbf x}|{\mathbf y})$ is approximated by an unnormalized Gaussian form, given by
\begin{align}
q({\bf{x}}) &\propto {\cal N}\left( {{\bf{y}};{{\bf{H}}_{{\rm{AF}}}}{\bf{x}},{\sigma_w ^2}{{\bf{I}}_{2N}}} \right)\prod\limits_{i = 1}^{2N} {e^ { - \frac{1}{2}{\lambda _i}x_i^2 + {\gamma _i}{x_i}}},
\end{align}
where $\lambda_i>0$ and $\gamma_i$ denote the parameters updated in each iteration. The mean $\boldsymbol{\mu}_{\rm{AF}}$ and variance $\boldsymbol{\Sigma}_{\rm{AF}}$ of the Gaussian approximation $q({\bf{x}})$ can be respectively calculated as
\begin{align}\label{eq:EPAF_PostMean}
\boldsymbol{\mu}_{\rm{AF}} = \boldsymbol{\Sigma}_{\rm{AF}}\left(\sigma_w^{-2}  {\bf{H}}_{{\rm{AF}}}^T\boldsymbol{y} + \boldsymbol{\gamma}\right),
\end{align}
\begin{align}\label{eq:EPAF_PostVar}
\boldsymbol{\Sigma}_{\rm{AF}} =\boldsymbol{\Psi}_{\mathrm{AF}}^{-1}= \left( \sigma_w^{-2} {\bf{H}}_{{\rm{AF}}}^T {\bf{H}}_{{\rm{AF}}} + \operatorname{diag}({\boldsymbol \lambda })\right)^{-1},
\end{align}
where $\boldsymbol{\Psi}_{\mathrm{AF}}=\sigma_w^{-2} {\bf{H}}_{{\rm{AF}}}^T {\bf{H}}_{{\rm{AF}}} + \operatorname{diag}({\boldsymbol \lambda })$ denotes the AF domain equalization matrix, \(\boldsymbol{\lambda} = [\lambda_1, \lambda_2, \ldots, \lambda_{2N}]^T\) and \(\boldsymbol{\gamma} = [\gamma_1, \gamma_2, \ldots, \gamma_{2N}]^T\). The marginal distribution of \( q(\mathbf{x}) \) for the $i$-th variable is denoted by \(q(x_i) = \mathcal{N}(x_i; \mu_i, \sigma_i^2)\), where \(\mu_i\) and \(\sigma_i^2\) are the \(i\)-th elements of \(\boldsymbol{\mu}_{\rm{AF}}\) and \(\mathrm{diag}(\boldsymbol{\Sigma}_{\rm{AF}})\), respectively. The true posterior marginal \(f(x_i)\) is approximated by \(q(x_i)\) under the assumed Gaussian prior approximation.

\subsubsection{Cavity Distribution Calculation}
For each variable $x_i$, the corresponding cavity distribution is defined as 
\begin{align}\label{eq:EPAF_CavityDis}
q^{\setminus i}(x_i) = \frac{q(x_i)}{e^ { - \frac{1}{2}{\lambda _i}x_i^2 + {\gamma _i}{x_i}}} \propto \mathcal{N}(x_i; \eta_i, \zeta_i^2),i \in \mathcal{I}_{2N},
\end{align}
where $\mathcal{I}_{2N} =\{1,2,\ldots,2N\}$, $\zeta_i^2 = \frac{\sigma_i^2}{1 - \sigma_i^2 \lambda_i}$, and $\eta_i = \zeta_i^2 \left( \frac{\mu_i}{\sigma_i^2} - \gamma_i \right)$.

\subsubsection{Computation of the True Posterior Marginal Distribution}
Based on the cavity distribution, the true posterior marginal distribution of each variable \(x_i\) is given by
\begin{align}\label{eq:EPAF_TruePost}
f(x_i) \propto q^{\setminus i}(x_i)  \mathbb{I}_{x_i \in \mathcal{A}}, i \in \mathcal{I}_{2N},
\end{align}
and its corresponding mean and variance are computed as \(\kappa_i = \mathbb{E}_{f(x_i)}[x_i]\) and \(\chi_i^2 = \mathbb{E}_{f(x_i)}[(x_i - \kappa_i)^2]\), respectively. Here, \(\mathbb{E}_{f(x_i)}[\cdot]\) denotes the expectation with respect to the posterior marginal distribution \(f(x_i)\).

\subsubsection{Parameter Update}
In the $(t+1)$-th iteration, the prior parameters \(\lambda_i\) and \(\gamma_i\) are updated to refine the Gaussian approximation via the moment matching method, yielding
\begin{align}\label{eq:EPAF_ParaUpdate1}
\lambda_i[t + 1] = \Delta \left( \frac{1}{\chi_i^2[t]} - \frac{1}{{\zeta}_i^2[t]} \right) + (1 - \Delta)\lambda_i[t],
\end{align}
and
\begin{align}\label{eq:EPAF_ParaUpdate2}
\gamma_i[t + 1] = \Delta \left( \frac{\kappa_i[t]}{\chi_i^2[t]} - \frac{\eta_i[t]}{{\zeta}_i^2[t]} \right) + (1 - \Delta)\gamma_i[t],
\end{align}
for $i \in \mathcal{I}_{2N}$, where $\Delta \in (0, 1]$ denotes the damping factor.
 

\begin{figure}
    \centering
\includegraphics[width=0.8\linewidth] {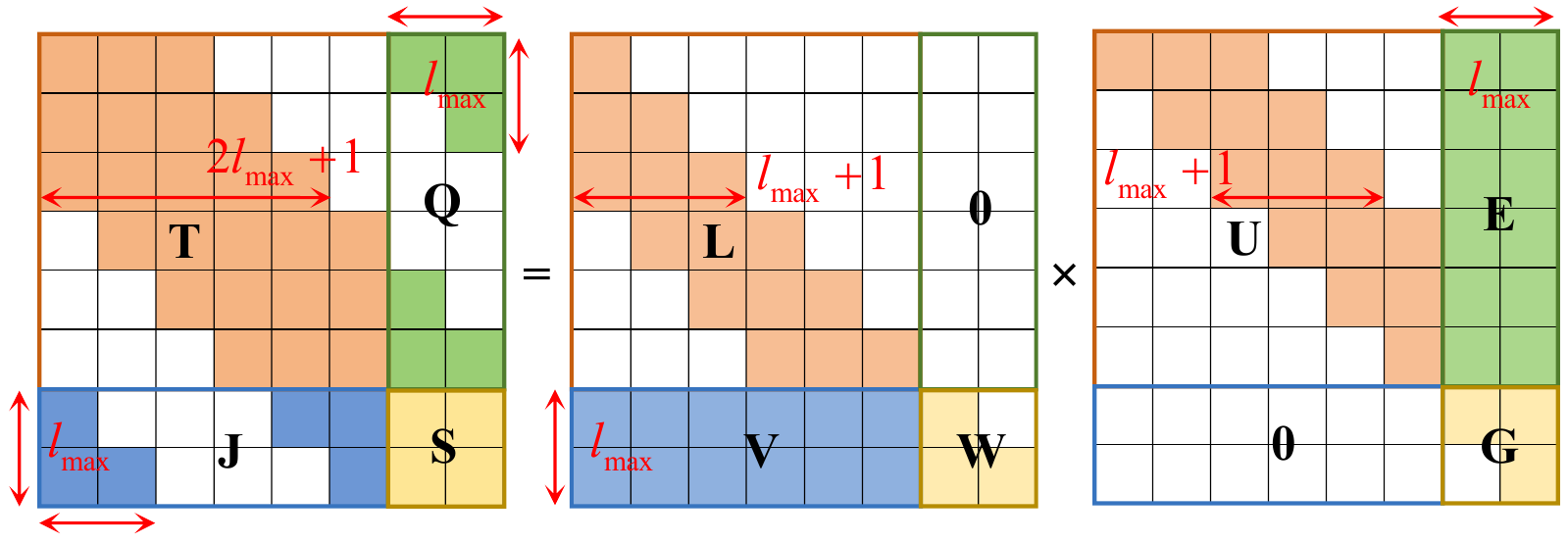}
    \caption[left]{The LU factorization of the quasi-banded matrix $\mathbf{B}$ with $N=8$ and $l_\mathrm{max}=2$.}
    \label{fig:LUDecomp}
    \vspace{-2em}
\end{figure}

\subsubsection{Low-Complexity Quasi-Banded Matrix Inversion for Posterior Covariance Computation} The main computational complexity during each EP-AF iteration arises from computing the posterior covariance matrix in \eqref{eq:EPAF_PostVar}, which requires a matrix inversion with complexity \(\mathcal{O}(N^3)\). To address this, we introduce a low-complexity matrix inversion method that leverages the quasi-banded structure of the AF domain channel matrix $\bar{\mathbf{H}}_{\mathrm{AF}}$ to reduce the computational complexity of \(\boldsymbol{\Psi}_{\rm{AF}}^{-1}\).


Based on \eqref{eq:Heff_Real} and \eqref{eq:EPAF_PostVar}, \(\boldsymbol{\Psi}_{\rm{AF}}\) is expressed in block form as
\begin{align}\label{eq:Fai_AF}
\boldsymbol{\Psi}_{\rm{AF}} = \begin{bmatrix} \mathbf{B} & -\mathbf{C} \\ \mathbf{C} & \mathbf{B} + \mathbf{D} \end{bmatrix},
\end{align}
where $\mathbf{B} = \sigma_w^{-2}\Re\left\{\mathbf{\Phi}_{\mathrm{AF}}\right\}+\operatorname{diag}(\hat{\boldsymbol \lambda })$, $\mathbf{C} = \sigma_w^{-2}\Im\left\{\mathbf{\Phi}_{\mathrm{AF}}\right\}$,   $\mathbf{D} = \operatorname{diag}(\check{\boldsymbol \lambda })-\operatorname{diag}(\hat{\boldsymbol \lambda })$, and \( \hat{\boldsymbol \lambda }, \check{\boldsymbol \lambda } \in \mathbb{R}^{N}  \) represent the first and last \( N \) entries of ${\boldsymbol{\lambda}}$, respectively. According to the structural analysis in Subsection~\ref{Sec:ChannelSpars}, both $\mathbf{B}$ and $\mathbf{C}$ are quasi-banded matrices with a bandwidth of
$ 2l_{\max} + 1$. Therefore, \(\boldsymbol{\Psi}_{\rm{AF}}^{-1}\) can be efficiently calculated via the block inversion formula as
\begin{align}\label{eq:Fai_AF_Inver}
\boldsymbol{\Psi}_{\rm{AF}}^{-1}=\left[\begin{array}{cc}
\mathbf{B}^{-1}-\mathbf{B}^{-1} \mathbf{C} \mathbf{\mathbf{P}}^{-1} \mathbf{C} \mathbf{B}^{-1} & \mathbf{B}^{-1} \mathbf{C} \mathbf{\mathbf{P}}^{-1} \\
-\mathbf{\mathbf{P}}^{-1} \mathbf{C} \mathbf{B}^{-1} & \mathbf{\mathbf{P}}^{-1}
\end{array}\right],
\end{align}
where $\mathbf{\mathbf{P}}=\mathbf{B}+\mathbf{D}+\mathbf{C} \mathbf{B}^{-1} \mathbf{C}$. 
LU factorization \cite{LU_Decomp} is employed to compute the inverse of the quasi-banded matrix \(\mathbf{B}\), whose decomposition is depicted in Fig.~\ref{fig:LUDecomp} and can be written as
\begin{align}
&{\bf{B }}=\left[ {\begin{array}{*{20}{l}}
{{{\bf{T}}_{\theta  \times \theta }}}&{{{\bf{Q}}_{\theta  \times {l_{\max }}}}}\\
{{{\bf{J}}_{{l_{\max }} \times \theta }}}&{{{\bf{S}}_{{l_{\max }} \times {l_{\max }}}}}
\end{array}} \right] \nonumber \\
&=
\left[ {\begin{array}{*{20}{l}}
{{{\bf{L}}_{\theta  \times \theta }}}&{{{\bf{0}}_{\theta  \times {l_{\max }}}}}\\
{{{\bf{V}}_{{l_{\max }} \times \theta }}}&{{{\bf{W}}_{{l_{\max }} \times {l_{\max }}}}}
\end{array}} \right]\left[ {\begin{array}{*{20}{l}}
{{{\bf{U}}_{\theta  \times \theta }}}&{{{\bf{E}}_{\theta  \times {l_{\max }}}}}\\
{{{\bf{0}}_{{l_{\max }} \times \theta }}}&{{{\bf{G}}_{{l_{\max }} \times {l_{\max }}}}}
\end{array}} \right],
\end{align}
where $\theta = N-l_{\max }$. By applying the block matrix inversion formula, \({{\bf{B}}^{ - 1}}\) can be equivalently expressed as
\begin{align}\label{eq:EPAF_InverB}
{{\bf{B}}^{ - 1}}= {\left[ {\begin{array}{*{20}{l}}
{\bf{U}}&{\bf{E}}\\
{\bf{0}}&{\bf{G}}
\end{array}} \right]^{ - 1}}{\left[ {\begin{array}{*{20}{l}}
{\bf{L}}&{\bf{0}}\\
{\bf{V}}&{\bf{W}}
\end{array}} \right]^{ - 1}} 
 \triangleq  \left[ {\begin{array}{*{20}{l}}
{{{\bf{B}}_{11}}}+{{{\bf{B}}_{12}}}&{{{\bf{B}}_2}}\\
{{{\bf{B}}_3}}&{{{\bf{B}}_4}}
\end{array}} \right]
\end{align}
where ${{\bf{B}}_{11}} = {{\bf{U}}^{{{ - 1}}}}{{\bf{L}}^{{{ - 1}}}}$, ${{\bf{B}}_{12}} = {{\bf{U}}^{{{ - 1}}}}{\bf{E}}{{\bf{B}}_4}{\bf{V}}{{\bf{L}}^{{{ - 1}}}}$, ${{\bf{B}}_2} =  - {{\bf{U}}^{{{ - 1}}}}{\bf{E}}{{\bf{B}}_4}$, ${{\bf{B}}_3} =- {{\bf{B}}_4}{\bf{V}}{{\bf{L}}^{{{ - 1}}}}$, and ${{\bf{B}}_4} = {{\bf{G}}^{{{ - 1}}}}{{\bf{W}}^{{{ - 1}}}}$.  Since \(l_{\max} \ll N\), the submatrices \(\mathbf{G}^{-1}\) and \(\mathbf{W}^{-1}\) in \(\mathbf{B}_4\) are small and can be directly inverted with complexity \(\mathcal{O}(l_{\max}^3)\). As \(\mathbf{L}\) and \(\mathbf{U}\) are lower and upper triangular banded matrices, their inverses \(\mathbf{L}^{-1}\) and \(\mathbf{U}^{-1}\) can be efficiently computed using forward and backward substitution with complexity \(\mathcal{O}(l_{\max}^2 N)\) \cite{LU_Decomp}. To further reduce the complexity of calculating \(\mathbf{B}_{11} = \mathbf{U}^{-1} \mathbf{L}^{-1}\), both \(\mathbf{U}^{-1}\) and \(\mathbf{L}^{-1}\) are approximated as banded matrices with bandwidth \(\phi\), where \(\phi  = kl_{\max}\) controls the accuracy of the inversion approximation. In addition, we observe that the non-zero elements of \(\mathbf{U}^{-1} \mathbf{E}\) are mainly concentrated in the first and last \(\phi\) rows, while those of \(\mathbf{V} \mathbf{L}^{-1}\) are concentrated in the first and last  \(\phi\) columns. Therefore, the matrices \(\mathbf{B}_{12}\), \(\mathbf{B}_2\), and \(\mathbf{B}_3\) can be efficiently computed using sparse matrix multiplications. After constructing \(\mathbf{B}^{-1}\), a similar low-complexity approach is applied to approximate the inversion of quasi-banded matrix \(\mathbf{P}\). Finally, we can obtain \(\boldsymbol{\Psi}_{\rm{AF}}^{-1}\) by substituting the approximated \(\mathbf{B}^{-1}\) and \(\mathbf{P}^{-1}\) into \eqref{eq:Fai_AF_Inver}.

\subsection{EP-T Detection for Doubly-Selective Channels}\label{Sec:EP-T}
The proposed low-complexity EP-T algorithm leverages the quasi-banded structure of the time domain channel matrix $\bar{\mathbf{H}}_\mathrm{T}$ to reduce the computational complexity associated with equalization matrix inversion. It approximates the posterior distribution of the time domain symbols as a Gaussian distribution and leverages the AF domain symbol constellation constraints to compute the posterior statistics.


Similar to \eqref{eq:RxSignal_AFD_Real}, the time domain complex-valued system model \eqref{eq:RxSym_TD} can be equivalently written in real-valued form as $\mathbf{r} = \mathbf{H}_{\mathrm{T}} \mathbf{s} + \mathbf{w}_{\mathrm{T}}$, where \(\mathbf{r}, \mathbf{s}, \mathbf{w}_{\mathrm{T}} \in \mathbb{R}^{2N}\) denote received signal, transmitted signal, and noise vectors in real-valued form, respectively, and \(\mathbf{H}_{\mathrm{T}} \in \mathbb{R}^{2N \times 2N}\) represents the real-valued equivalent channel matrix in the time domain. The main processes of the proposed EP-T algorithm are detailed below:
\subsubsection{Posterior Approximation in the Time Domain}
With the Gaussian approximation applied to the time domain symbols, the posterior probability of 
$\mathbf{s}$ is expressed as
\begin{align}
q({\bf{s}}) &\propto {\cal N}\left( {{\bf{r}};{{\bf{H}}_{{\rm{T}}}}{\bf{s}},{\sigma_w ^2}{{\bf{I}}_{2N}}} \right)\prod\limits_{i = 1}^{2N} {e^ { - \frac{1}{2}\tilde{\lambda _i}s_i^2 + \tilde{\gamma _i}{s_i}}},
\end{align}
where $\tilde{\lambda _i}>0$ and $\tilde{\gamma _i}$ are the parameters updated per iteration. The mean $\boldsymbol{\mu}_{\rm{T}}$ and variance $\boldsymbol{\Sigma}_{\rm{T}}$ of $q(\mathbf{s})$ are computed as 
\begin{align}\label{eq:EPT_PostMean}
\boldsymbol{\mu}_{\rm{T}} = \boldsymbol{\Sigma}_\mathrm{T}\left(\sigma_w^{-2}  {\bf{H}}_{{\rm{T}}}^T\mathbf{r} + \tilde{\boldsymbol{\gamma}}\right),
\end{align}
\begin{align}\label{eq:EPT_PostVar}
\boldsymbol{\Sigma}_{\rm{T}} =\boldsymbol{\Psi}_{\mathrm{T}}^{-1}= \left( \sigma_w^{-2} {\bf{H}}_{{\rm{T}}}^T {\bf{H}}_{{\rm{T}}} + \operatorname{diag}(\tilde{\boldsymbol{\lambda}})\right)^{-1},
\end{align}
where $\boldsymbol{\Psi}_{\mathrm{T}}=\sigma_w^{-2} {\bf{H}}_{{\rm{T}}}^T {\bf{H}}_{{\rm{T}}} + \operatorname{diag}(\tilde{\boldsymbol{\lambda}})$ is the time domain equalization matrix.
Accordingly, the marginal distribution of the $i$-th variable is given by $q(s_i) = \mathcal{N}(s_i; \tilde{\mu}_i, \tilde{\sigma}_i^2)$, where $\tilde{\mu}_i$ and $\tilde{\sigma}_i^2$ denote the $i$-th entries of $\boldsymbol{\mu}_{\mathrm{T}}$ and the diagonal of $\boldsymbol{\Sigma}_{\mathrm{T}}$, respectively.

As illustrated in Figs.~\ref{fig:H_vmax0} and \ref{fig:H_vmax1}, both the time domain Gram matrix $\mathbf{\Phi}_{\mathrm{T}}$ under doubly-selective channels and the AF domain Gram matrix $\mathbf{\Phi}_{\mathrm{AF}}$ under frequency-selective channels exhibit an identical quasi-banded sparse structure with a bandwidth of \(2l_{\max} + 1\). Consequently, the corresponding equalization matrices $\boldsymbol{\Psi}_{\mathrm{T}}$ and $\boldsymbol{\Psi}_{\mathrm{AF}}$ have the same sparsity pattern. Therefore, $\boldsymbol{\Psi}_{\mathrm{T}}^{-1}$ can be efficiently computed using the low-complexity quasi-banded matrix inversion method presented in Subsection~\ref{Sec:EP-AF}. 

\subsubsection{Cavity Distribution Calculation in the Time Domain}
Referring to \eqref{eq:EPAF_CavityDis}, the cavity marginal distribution for each variable $s_i$ is given by $q^{\setminus i} (s_i) = \frac{q(s_i)}{e^{ - \frac{1}{2} \tilde{\lambda}_i s_i^2+\tilde{\gamma}_i s_i}} \propto \mathcal{N} \left( s_i ; \tilde{\eta}_i , \tilde{\zeta}_i^2 \right) ,i \in \mathcal{I}_{2N}$,
 where ${\tilde \zeta}_i^2 = \frac{{\tilde \sigma}_i^2}{1 - {\tilde \sigma}_i^2 \tilde{\lambda}_i}$ and $\eta_i = {\tilde \zeta}_i^2 \left( \frac{{\tilde \mu}_i}{{\tilde \sigma}_i^2} - {\tilde \gamma}_i \right)$.  Define the time domain cavity mean and variance vectors as $\boldsymbol{\eta}_{\mathrm{T}} = \left[\tilde{\eta}_1, \tilde{\eta}_2, \ldots, \tilde{\eta}_{2N}\right]^T=\left [\hat{\boldsymbol{\eta}}_{\mathrm{T}}^{T}, \check{\boldsymbol{\eta}}_{\mathrm{T}}^{T}\right]^{T}$, 
$\boldsymbol{\zeta}_{\mathrm{T}} = \left[\tilde{\zeta}_1^2, \tilde{\zeta}_2^2, \ldots, \tilde{\zeta}_{2N}^2\right]^T=\left[\hat{\boldsymbol{\zeta}}_{\mathrm{T}}^{T}, \check{\boldsymbol{\zeta}}_{\mathrm{T}}^{T}\right]^{T}$, where  \( \hat{\boldsymbol{\eta}}_{\mathrm{T}}, \check{\boldsymbol{\eta}}_{\mathrm{T}},\hat{\boldsymbol{\zeta}}_{\mathrm{T}}, \check{\boldsymbol{\zeta}}_{\mathrm{T}} \in \mathbb{R}^{N} \) correspond to the first and last $N$ elements of the respective vectors.

\subsubsection{Transformation of the Cavity Distribution into the AF Domain}
The cavity distribution is transformed from the time domain to the AF domain via the unitary transformation, and the corresponding mean and variance vectors are given by
\begin{align}\label{eq:EPT_Cavity_AFMean1}
 \hat{\boldsymbol{\eta}}_{\mathrm{AF}} = \Re \{ \mathbf{A} \hat{\boldsymbol{\eta}}_{\mathrm{T}} \} - \Im \{ \mathbf{A} \check{\boldsymbol{\eta}}_{\mathrm{T}} \}, 
 \end{align}
\begin{align}\label{eq:EPT_Cavity_AFMean2}
 \check{\boldsymbol{\eta}}_{\mathrm{AF}}& = \Im \{ \mathbf{A} \hat{\boldsymbol{\eta}}_{\mathrm{T}} \} + \Re \{ \mathbf{A} \check{\boldsymbol{\eta}}_{\mathrm{T}}\},
\end{align}
and
 \begin{align}\label{eq:EPT_Cavity_AFVar1}
 \hat{\boldsymbol{\zeta}}_{\mathrm{AF}} = {\mathrm{diag}}\left( \mathbf{A} \operatorname{diag} ( \hat{\boldsymbol{\zeta}}_{\mathrm{T}} ) \mathbf{A}^H \right),
\end{align}
 \begin{align}\label{eq:EPT_Cavity_AFVar2}
 \check{\boldsymbol{\zeta}}_{\mathrm{AF}} =
 {\mathrm{diag}}\left( \mathbf{A} \operatorname{diag} ( \check{\boldsymbol{\zeta}}_{\mathrm{T}} ) \mathbf{A}^H \right).
\end{align}
The complete AF domain cavity mean and variance vectors are then formed as \( \boldsymbol{\eta}_{\mathrm{AF}} = \left[\hat{\boldsymbol{\eta}}_{\mathrm{AF}}^{T}, \check{\boldsymbol{\eta}}_{\mathrm{AF}}^{T}\right]^{T} \) and \( \boldsymbol{\zeta}_{\mathrm{AF}} = \left[\hat{\boldsymbol{\zeta}}_{\mathrm{AF}}^{T}, \check{\boldsymbol{\zeta}}_{\mathrm{AF}}^{T}\right]^{T} \), respectively. 

\subsubsection{Computation of the True Posterior Marginal Distribution}
Based on \eqref{eq:EPAF_TruePost} and \eqref{eq:EPT_Cavity_AFMean1}-\eqref{eq:EPT_Cavity_AFVar2}, the mean vector \( {\boldsymbol{\kappa}}_{\mathrm{AF}} \) and variance vector \( \boldsymbol{\chi}_{\mathrm{AF}} \) of the true posterior marginal distribution $f(\mathbf{x})$ can be computed. These vectors are then partitioned as \( {\boldsymbol{\kappa}}_\mathrm{AF} = \left[\hat{\boldsymbol{\kappa}}_\mathrm{AF}^T, \check{\boldsymbol{\kappa}}_\mathrm{AF}^T\right]^T \), \( \boldsymbol{\chi}_\mathrm{AF} = \left[\hat{\boldsymbol{\chi}}_\mathrm{AF}^T, \check{\boldsymbol{\chi}}_\mathrm{AF}^T\right]^T \), where \( \hat{\boldsymbol{\kappa}}_{\mathrm{AF}}, \check{\boldsymbol{\kappa}}_{\mathrm{AF}} \in \mathbb{R}^{N}  \) and \( \hat{\boldsymbol{\chi}}_{\mathrm{AF}}, \check{\boldsymbol{\chi}}_{\mathrm{AF}} \in \mathbb{R}^{N}  \) denote the first and last \( N \) entries of ${\boldsymbol{\kappa}}_\mathrm{AF}$ and $\boldsymbol{\chi}_\mathrm{AF}$, respectively.

\subsubsection{Transformation of the Posterior Distribution into the Time domain}
The posterior marginal distribution is mapped from the AF domain to the time domain, yielding
$\hat{\boldsymbol{\kappa}}_{\mathrm{T}} = \Re \{ \mathbf{A}^H \hat{\boldsymbol{\kappa}}_{\mathrm{AF}} \} - \Im \{ \mathbf{A}^H \check{\boldsymbol{\kappa}}_{\mathrm{AF}} \}$,  
$\check{\boldsymbol{\kappa}}_{\mathrm{T}} = \Im \{ \mathbf{A}^H \hat{\boldsymbol{\kappa}}_{\mathrm{AF}} \} + \Re \{ \mathbf{A}^H \check{\boldsymbol{\kappa}}_{\mathrm{AF}}\}$, 
$\hat{\boldsymbol{\chi}}_{\mathrm{T}} = {\mathrm{diag}}\left( \mathbf{A}^H \operatorname{diag} ( \hat{\boldsymbol{\chi}}_{\mathrm{AF}} ) \mathbf{A} \right)$,
 and $\check{\boldsymbol{\chi}}_{\mathrm{T}} =
 {\mathrm{diag}}\left( \mathbf{A}^H \operatorname{diag} ( \check{\boldsymbol{\chi}}_{\mathrm{AF}} ) \mathbf{A} \right)$.
The full time domain posterior mean and variance vectors are then constructed as \( {\boldsymbol{\kappa}}_{\mathrm{T}} = [\hat{\boldsymbol{\kappa}}_{\mathrm{T}}^{T}, \check{\boldsymbol{\kappa}}_{\mathrm{T}}^{T}]^{T}=[\tilde{\kappa}_1, \tilde{\kappa}_2, \ldots, \tilde{\kappa}_{2N}]^T \) and \( \boldsymbol{\chi}_{\mathrm{T}} = [\hat{\boldsymbol{\chi}}_{\mathrm{T}}^{T}, \check{\boldsymbol{\chi}}_{\mathrm{T}}^{T}]^{T}=[\tilde{\chi}_1^2, \tilde{\chi}_2^2, \ldots, \tilde{\chi}_{2N}^2]^T \), where \( \hat{\boldsymbol{\kappa}}_{\mathrm{T}}, \check{\boldsymbol{\kappa}}_{\mathrm{T}} \in \mathbb{R}^{N}  \) and \( \hat{\boldsymbol{\chi}}_{\mathrm{T}}, \check{\boldsymbol{\chi}}_{\mathrm{T}} \in \mathbb{R}^{N}\) represent the first and last \( N \) elements of $\kappa_\text{T}$ and $\boldsymbol{\chi}_\text{T}$, respectively.

\subsubsection{Parameter Update in the Time Domain}
The time domain prior parameters $\tilde{\lambda}_i[t + 1]$ and $\tilde{\gamma}_i[t + 1]$ are updated in the same manner as \eqref{eq:EPAF_ParaUpdate1} and \eqref{eq:EPAF_ParaUpdate2} with  $ {\lambda}_i[t]$, ${{\chi}_i^2}[t]$, ${{\zeta}_i^2}[t]$, ${\kappa}_i[t]$, ${\gamma}_i[t]$, and ${{\eta}_i}[t]$ replaced by $\tilde {\lambda}_i[t]$, ${\tilde{\chi}_i^2}[t]$, ${\tilde{\zeta}_i^2}[t]$, $\tilde{\kappa}_i[t]$, $\tilde{\gamma}_i[t]$, and ${\tilde{\eta}_i}[t]$, respectively.

\subsection{Complexity Analysis}
The conventional EP-AF detector exhibits a complexity  of \( \mathcal{O}(N^3)\) in each iteration, resulting in a total complexity  of \( \mathcal{O}(N^3 T_{\mathrm{EP}}) \), where \( T_{\mathrm{EP}} \) denotes the number of iterations. Based on Subsection \ref{Sec:EP-AF}, the proposed low-complexity EP-AF detector reduces the per-iteration complexity  to \( \mathcal{O}(N k^2 l_{\max}^2) \) by employing the quasi-banded matrix inversion method, yielding a total complexity  of \( \mathcal{O}(N k^2 l_{\max}^2 T_{\mathrm{EP}}) \). Furthermore, the proposed low-complexity EP-T detector adopts the same inversion strategy and additionally performs unitary transformation in each iteration for domain information exchange with a complexity  of $\mathcal{O}(N \log N)$, leading to an overall complexity of $\mathcal{O}\left[ (N k^2 l_{\max}^2 + N \log N) T_{\mathrm{EP}} \right]$.


\section{Simulation Results}
This section presents simulation results to evaluate the performance of the proposed EP-AF and EP-T detectors in AFDM systems. The number of subcarriers is set to $N=256$, and the carrier frequency is $f_c$ = 4 GHz.
The maximum delay spread is $l_{\mathrm{max} }={P}-1$, and the channel coefficients $h_i$ follow a complex-valued Gaussian distribution, i.e., \(h_i \sim \mathcal{CN}(0, 1/{P})\). The maximum number of iterations is 10 for the EP-AF and EP-T detectors, and 50 for the MRC detector. For the proposed low-complexity EP-AF and EP-T detectors, a damping factor of $\Delta = 0.2$ is applied, and the band approximation parameter is set to $k=8$ to control the matrix inversion accuracy.

\begin{figure}
    \centering
    \includegraphics[width=65mm]{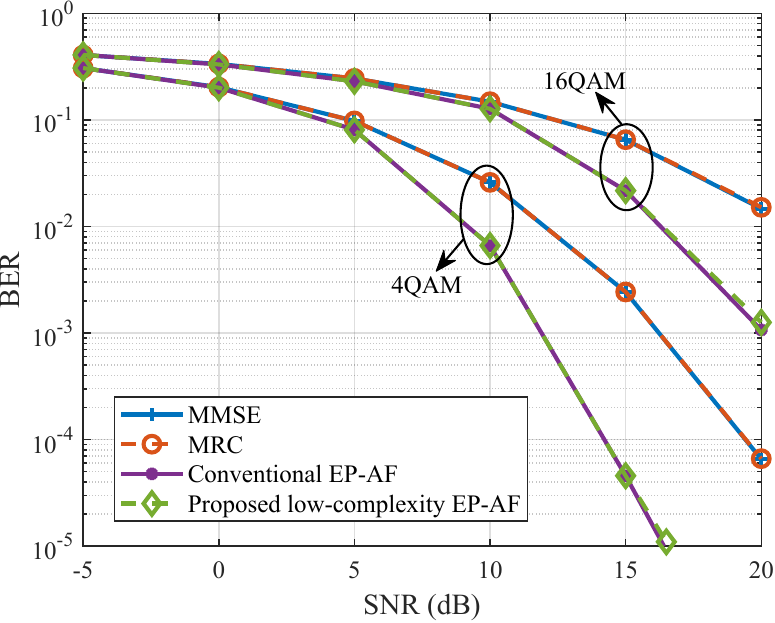}
    \caption{ 
    {BER performance of AFDM using MMSE, MRC, conventional EP-AF, and the proposed low-complexity EP-AF detectors under frequency-selective channels.}}
\label{fig:BER_LTI}
\vspace{-1.8em}
\end{figure}

Fig.~\ref{fig:BER_LTI} compares the BER performance of AFDM systems using MMSE, MRC, conventional EP-AF, and the proposed low-complexity EP-AF detectors. Both 4QAM and 16QAM modulation schemes are considered, with 9-path frequency-selective channels. It can be observed that the low-complexity EP-AF detector outperforms both the MMSE and MRC detectors. Specifically, at a BER of $10^{-4}$, it achieves a signal-to-noise ratio (SNR) gain of approximately 5.2~dB compared to the MMSE detector under 4QAM. In addition, the proposed low-complexity EP-AF detector with the low-complexity quasi-banded matrix inversion approximation is nearly the same as that of the conventional EP-AF detector without the approximation calculation.

\begin{figure}
    \centering
    \includegraphics[width=65mm]{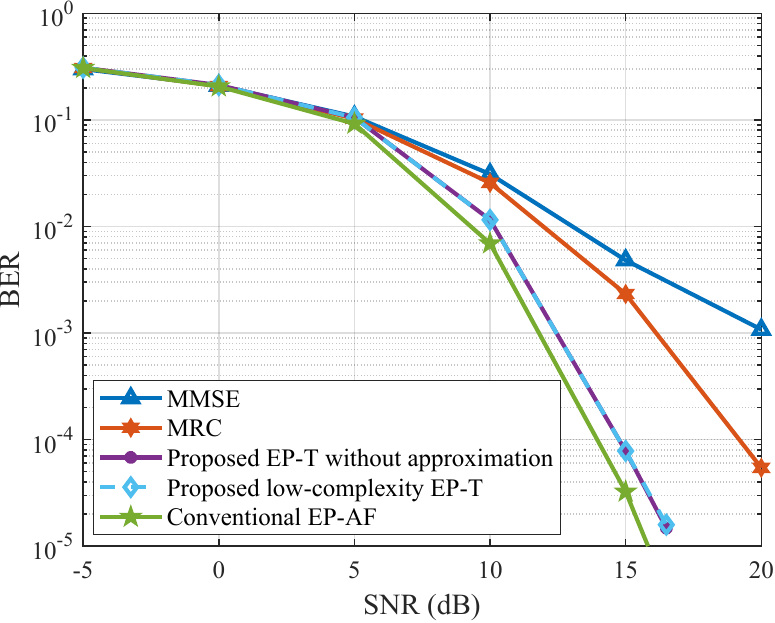}
    \caption{BER performance of AFDM using MMSE, MRC, conventional EP-AF, the proposed EP-T without the approximation calculation, and the proposed low-complexity EP-T detectors under doubly-selective channels.}
    \label{fig:BER_LTVDiffDet}
    \vspace{-1.8em}
\end{figure}

Fig.~\ref{fig:BER_LTVDiffDet} illustrates the BER performance of AFDM systems employing MMSE, MRC, conventional EP-AF, the proposed EP-T without the matrix inversion approximation, and the low-complexity EP-T detectors for 4QAM. An 8-path doubly selective channel is considered, with the normalized maximum Doppler shift set to $\nu_{\max} = 1$ \cite{AFDM}.
The Doppler shift for each path is generated according to the Jakes model, $\nu_i = \nu_{\mathrm{max}} \cos(\omega_i)$, with $\omega_i$ uniformly distributed over $[-\pi, \pi)$. It can be observed that the proposed EP-T detector achieves significantly better BER performance than both MMSE and MRC detectors. Specifically, it achieves an SNR gain of approximately 7.8~dB compared to the MRC detector at a BER of $10^{-3}$, and yields about 4.4~dB gain over the MMSE detector at a BER of $10^{-4}$. In addition, the proposed low-complexity EP-T detector attains BER performance nearly identical to that without matrix inversion approximation. The performance gap between the proposed low-complexity EP-T detector and the conventional EP-AF detector can be attributed to the fact that the former updates the prior parameters of the time domain symbols, whose set is significantly larger than that of the AF domain symbols used in the latter.



\begin{figure}
    \centering
    \includegraphics[width=65mm]{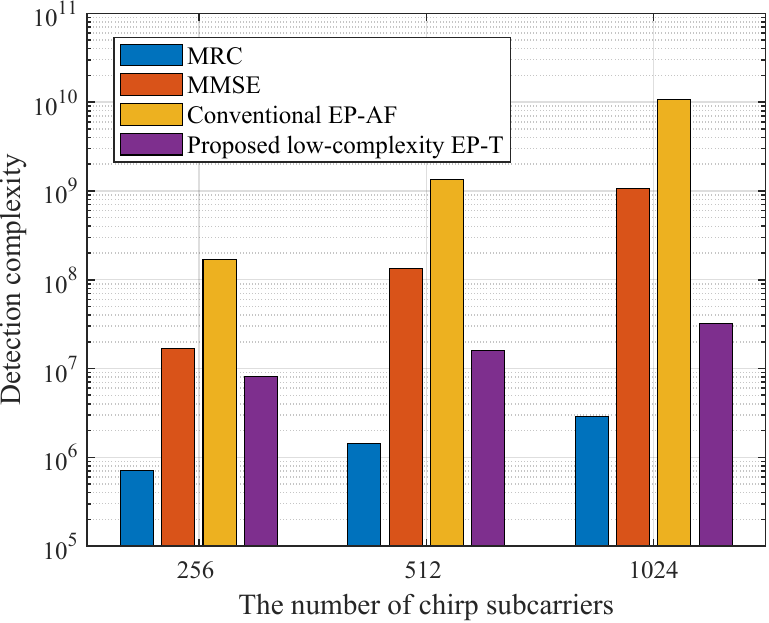}
    \caption{Complexity comparison of AFDM using  MMSE, MRC, conventional EP-AF, and the proposed low-complexity EP-T detectors under doubly-selective channels.}
   \label{fig:Complexity_DiffDetection}
   \vspace{-1.8em}
\end{figure}

Fig.~\ref{fig:Complexity_DiffDetection} compares the complexity of the MMSE, MRC, conventional EP-AF, and the proposed low-complexity EP-T detectors under doubly-selective channels, measured in terms of the number of multiplications. 
It can be seen that the complexity of the proposed EP-T detector is nearly one order of magnitude lower than that of the MMSE detector, and approximately two orders of magnitude lower than that of the conventional EP-AF detector. Furthermore, compared to the MRC detector, the proposed detector incurs about one order of magnitude higher complexity. Although the MRC detector offers lower complexity, it suffers from significantly degraded BER performance compared with the proposed EP-T detector. Therefore, the proposed low-complexity EP-T detector achieves a favorable trade-off between detection performance and complexity.

\vspace{-0.5em}
\section{Conclusion}
In this paper, we have proposed a dual-domain low-complexity EP detection framework for AFDM systems. By fully exploiting the quasi-banded sparsity of the effective channel matrices in both the time and AF domains, we have developed a low-complexity EP-AF detector and EP-T detector designed for frequency-selective channels and doubly-selective channels, respectively. The proposed EP-T detector avoids large-scale matrix inversion in the AF domain by transferring the Gaussian approximate inference to the time domain, thereby reducing the overall detection complexity from cubic to log-linear order. Simulation results have shown that the proposed detectors are attractive for future high mobility communication systems by exploiting the sparsity of effective channel matrices in different domains.

\end{document}